\documentclass[conference]{IEEEtran}
\IEEEoverridecommandlockouts

\usepackage{cite}
\usepackage{amsmath,amssymb,amsfonts}
\usepackage{graphicx}
\usepackage{textcomp}
\usepackage{xcolor}
\usepackage{algorithm}
\usepackage{algpseudocode}
\usepackage{comment}
\usepackage{url}
\usepackage{booktabs}
\usepackage{multirow}
\usepackage{hyperref}
\usepackage{orcidlink}

\def\BibTeX{{\rm B\kern-.05em{\sc i\kern-.025em b}\kern-.08em
    T\kern-.1667em\lower.7ex\hbox{E}\kern-.125emX}}

\makeatletter
\newcommand{\linebreakand}{
  \end{@IEEEauthorhalign}
  \hfill\mbox{}\par
  \mbox{}\hfill\begin{@IEEEauthorhalign}
}
\makeatother

\makeatletter
\newenvironment{breakablealgorithm}
  {
   \begin{flushleft}
     \refstepcounter{algorithm}
     \hrule height.8pt depth0pt \kern2pt
     \renewcommand{\caption}[2][\relax]{
       {\raggedright\textbf{\fname@algorithm~\thealgorithm} ##2\par}%
       \ifx\relax##1\relax 
         \addcontentsline{loa}{algorithm}{\protect\numberline{\thealgorithm}##2}%
       \else 
         \addcontentsline{loa}{algorithm}{\protect\numberline{\thealgorithm}##1}%
       \fi
       \kern2pt\hrule\kern2pt
     }
  }{
     \kern2pt\hrule\relax
   \end{flushleft}
  }
\makeatother

\begin{document}

\title{Implementing Large Quantum Boltzmann Machines as Generative AI Models for Dataset Balancing
}

\author{\IEEEauthorblockN{Salvatore Sinno \orcidlink{0009-0002-9177-5161}\IEEEauthorrefmark{1}\IEEEauthorrefmark{2}, Markus Bertl \orcidlink{0000-0003-0644-8095}\IEEEauthorrefmark{1}\IEEEauthorrefmark{3}, Arati Sahoo \orcidlink{0009-0007-3143-0344}\IEEEauthorrefmark{1}, Bhavika Bhalgamiya \orcidlink{0009-0001-8586-4531}\IEEEauthorrefmark{1},\\
Thomas Gro{\ss} \orcidlink{0000-0002-7766-2454
 }\IEEEauthorrefmark{2} and Nicholas Chancellor \orcidlink{0000-0002-1293-0761}\IEEEauthorrefmark{2}}\\

\IEEEauthorblockA{\IEEEauthorrefmark{1}NextGen Computing Research Group, Unisys, Blue Bell, Pennsylvania, USA\\
Email: \{salvatore.sinno, markus.bertl, arati.sahoo, bhavika.bhalgamiya\}@unisys.com}
\IEEEauthorblockA{\IEEEauthorrefmark{2}School of Computing, Newcastle University, Newcastle upon Tyne, United Kingdom}
\IEEEauthorblockA{\IEEEauthorrefmark{3}Department of Information Systems and Operations Management, Vienna University of Economics and Business, Austria}
}

\maketitle

\begin{abstract}
This study explores the implementation of large Quantum Restricted Boltzmann Machines (QRBMs), a key advancement in Quantum Machine Learning (QML), as generative models on D-Wave's Pegasus quantum hardware to address dataset imbalance in Intrusion Detection Systems (IDS). By leveraging Pegasus's enhanced connectivity and computational capabilities, a QRBM with 120 visible and 120 hidden units was successfully embedded, surpassing the limitations of default embedding tools. The QRBM synthesized over 1.6 million attack samples, achieving a balanced dataset of over 4.2 million records. Comparative evaluations with traditional balancing methods, such as SMOTE and RandomOversampler, revealed that QRBMs produced higher-quality synthetic samples, significantly improving detection rates, precision, recall, and F\textsubscript{1} score across diverse classifiers. The study underscores the scalability and efficiency of QRBMs, completing balancing tasks in milliseconds. These findings highlight the transformative potential of QML and QRBMs as next-generation tools in data preprocessing, offering robust solutions for complex computational challenges in modern information systems.\\
\end{abstract}

\begin{IEEEkeywords}
Quantum Computing, Quantum Machine Learning (QML), Quantum Restricted Boltzmann Machines (QRBM), Restricted Boltzmann Machines (RBM), Dataset Balancing, Unbalanced Data, Generative Models, Intrusion Detection Systems.
\end{IEEEkeywords}

\section{Introduction}
Balanced datasets have long been recognized as essential for effective machine learning (ML), enabling models to generalize and produce robust predictions. In the context of intrusion detection systems (IDS), the importance of balanced datasets becomes even more pronounced. IDS are critical for securing networks by identifying malicious activities amidst normal traffic. However, real-world datasets used for training IDS have often suffered from severe class imbalance, where the volume of normal traffic significantly outweighed that of intrusion attempts. This imbalance has limited the effectiveness of ML models, resulting in biased predictions favouring the majority class, reduced detection rates for rare but critical intrusions, and increased false positive rates.

The challenges posed by class imbalance were multifaceted. ML models trained on imbalanced datasets tended to optimize for the majority class, neglecting the minority class. Traditional balancing methods, such as oversampling and undersampling or Synthetic Minority Oversampling Technique (SMOTE) \cite{chawla2002smote}, have significant drawbacks. Oversampling introduced redundancy and increased the risk of overfitting; undersampling often discarded valuable information from the majority class. SMOTE can not handle class boundaries effectively and has high-dimensional or sequential data challenges. These limitations underscored the need for novel approaches to address class imbalance in IDS datasets.

This study leveraged the potential of quantum computing to address these challenges, focusing specifically on quantum-restricted Boltzmann Machines (QRBMs). QRBMs, as generative models, can learn the statistical properties of datasets and synthesise realistic data samples. By utilising principles of quantum mechanics, QRBMs could efficiently explore high-dimensional data spaces, making them particularly suited for generating synthetic data to balance imbalanced datasets. Unlike classical methods, QRBMs exploited quantum superposition and entanglement to represent complex probability distributions more effectively, offering significant advantages in addressing class imbalance.

The results and contributions of the study are summarized as follows:

\begin{itemize}
    \item \textbf{Largest QRBM Implementation:} Successfully implemented a large QRBM (120 visible and 120 hidden units) on D-Wave’s Pegasus architecture, overcoming the limitations of D-Wave’s default embedding tools. This implementation highlights the potential of QRBMs to act as powerful generative models for balancing even the most complex datasets.
    
    \item \textbf{Generative Modeling for IDS:} Demonstrated the efficacy of QRBMs in generating high-quality synthetic data, achieving superior precision, recall, and F\textsubscript{1} scores compared to traditional methods. 

    \item \textbf{Improved IDS Performance:} Evaluated the impact of QRBM-generated datasets on IDS performance metrics. Results showed significant improvements in detection rates and reductions in false positives, demonstrating QRBM’s ability to enhance IDS reliability and robustness.

\end{itemize}

\section{Related Work}
\label{sec:related}
Class imbalance is a persistent challenge in ML, particularly in fields like IDS, where the normal traffic volume far exceeds intrusion events. Several classical methods have been proposed to address this imbalance, including SMOTE and Random Oversampling. SMOTE generates synthetic samples by interpolating between existing minority class samples, while Random Oversampling duplicates minority class samples to achieve balance \cite{chawla2002smote, he2008learning}. Although these methods improve dataset balance, they often lead to issues such as overfitting and redundancy, limiting their effectiveness in real-world scenarios \cite{garcia2012effectiveness} \cite{mohammed2020machine}.

Previous studies, such as those by Gopalan et al. \cite{gopalan2021balancing} and Abdulrahman et al. \cite{abdulrahman2020toward}, emphasize balancing techniques for IDS datasets, highlighting the challenges of achieving fair evaluation metrics. Abdulrahman et al. \cite{abdulrahman2020toward} investigate balancing the CICIDS2017 dataset using classical techniques, which, while effective to some extent, often fail to capture nuanced feature interactions within the data. Liu et al. \cite{liu2022data} propose using Generative Adversarial Networks (GANs) for data balancing, showcasing their potential in generating robust datasets for IDS tasks. 

QRBMs have emerged as a promising alternative to classical methods for handling imbalanced datasets. QRBMs, an extension of classical Restricted Boltzmann Machines (RBMs), leverage the principles of quantum computing to model complex probability distributions and generate synthetic data. Unlike classical RBMs, QRBMs utilize quantum annealing to efficiently sample from high-dimensional energy landscapes, making them particularly well-suited for generative tasks \cite{hinton2006reducing, amin2018quantum}. Prior research has demonstrated the potential of QRBMs in various machine-learning applications, including anomaly detection and classification tasks  \cite{amin2018quantum, dixit2023quantum}.

Implementing QRBMs on quantum hardware poses unique challenges, particularly concerning embedding large models on physical quantum processors. D-Wave’s Chimera topology, an early quantum annealing architecture, provided limited qubit connectivity, which constrained the scalability of QRBM implementations \cite{venturelli2015quantum}. The introduction of D-Wave’s Pegasus topology addressed these limitations by offering enhanced qubit connectivity and increased computational capacity \cite{dwave2020pegasus}. Recent studies have explored methods for minor embedding of problems on Pegasus, showcasing significant improvements in performance and scalability \cite{dattani2019pegasus,pelofske2023clique}.

The proposed algorithm demonstrates remarkable flexibility and efficiency in embedding RBMs on D-Wave's Pegasus architecture. It enables the minor embedding up to a 172x120 RBM by optimizing chain lengths to remain short while maximizing the utilization of qubits for visible and hidden nodes.

This algorithm outperforms D-Wave’s default embedding tool. The ability to efficiently embed larger RBMs underscores the algorithm’s potential to enhance quantum annealing applications, particularly in generative modelling and data balancing tasks.

\section{Restricted Boltzmann Machines}
\label{sec:rbm}
RBMs are stochastic neural networks that model the joint probability distribution of visible and hidden variables. They are energy-based models that learn to represent data by minimizing an energy function. RBMs consist of two layers: a visible layer $\mathbf{v}$ representing observed data and a hidden layer $\mathbf{h}$ that captures latent features. There are no intra-layer connections, simplifying the computation and making the model tractable \cite{hinton2006reducing}.

The energy of a configuration $(\mathbf{v}, \mathbf{h})$ of the visible and hidden units in an RBM is defined as:
\begin{equation}
E(\mathbf{v}, \mathbf{h}) = -\sum_{i \in \text{visible}} b_i v_i - \sum_{j \in \text{hidden}} c_j h_j - \sum_{i,j} v_i W_{ij} h_j,
\label{eq:energy}
\end{equation}
where $b_i$ and $c_j$ are biases for the visible and hidden units, respectively, and $W_{ij}$ represents the weights of the connection between visible unit $i$ and hidden unit $j$. This energy function determines the likelihood of a given configuration; lower energy corresponds to higher probability.

The joint probability of a visible vector $\mathbf{v}$ and a hidden vector $\mathbf{h}$ is expressed using the Boltzmann distribution:
\begin{equation}
P(\mathbf{v}, \mathbf{h}) = \frac{\exp(-E(\mathbf{v}, \mathbf{h}))}{Z},
\end{equation}
where $Z$ is the partition function defined as:
\begin{equation}
Z = \sum_{\mathbf{v}, \mathbf{h}} \exp(-E(\mathbf{v}, \mathbf{h})).
\end{equation}
The partition function normalizes the probability distribution, ensuring that $P(\mathbf{v}, \mathbf{h})$ is valid.

\subsection{Conditional Independence and Sampling}
One of the key features of RBMs is that the visible and hidden units are conditionally independent of each other. This allows efficient sampling from the distribution:
\begin{align}
P(h_j = 1 | \mathbf{v}) &= \sigma\left(c_j + \sum_{i} W_{ij} v_i\right), \\
P(v_i = 1 | \mathbf{h}) &= \sigma\left(b_i + \sum_{j} W_{ij} h_j\right),
\end{align}
where $\sigma(x)$ is the sigmoid function $\sigma(x) = 1 / (1 + e^{-x})$. This property enables Gibbs sampling for training and inference.

RBMs are trained by maximizing the likelihood of the observed data $\mathbf{v}$. Direct computation of the log-likelihood gradient is intractable due to the partition function $Z$. To overcome this, Contrastive Divergence (CD) is used as an approximation \cite{hinton2002training}. The update rule for the weights is given by:
\begin{equation}
\Delta W_{ij} = \langle v_i h_j \rangle_{\text{data}} - \langle v_i h_j \rangle_{\text{model}},
\end{equation}
where $\langle \cdot \rangle_{\text{data}}$ represents expectations under the data distribution, and $\langle \cdot \rangle_{\text{model}}$ represents expectations under the model distribution.

\subsection{RBMs as Generative Models}
RBMs can be used as generative models as they learn and replicate a dataset's statistical properties. Once trained, an RBM can generate new data samples by alternately sampling from the visible and hidden layers in a process known as Gibbs sampling \cite{casella1992explaining}. This iterative sampling process alternates between activating the hidden units based on the visible units and vice versa, gradually producing new visible configurations that align with the learned data distribution. 

This generative capability makes RBMs a powerful tool for several applications. One of their primary uses is in dataset augmentation, where new data samples are synthesized to address challenges like limited data availability or class imbalance. For example, in highly imbalanced datasets, an RBM can generate synthetic samples for under-represented classes, effectively increasing their prevalence and improving the performance of downstream supervised learning models \cite{hinton2012practical}. Unlike oversampling techniques that merely duplicate existing data points, RBMs generate entirely new samples that preserve the statistical diversity of the dataset, reducing the risk of overfitting.

\subsection{Quantum Extensions}
 
In quantum computing, the concept of generative modelling with RBMs has been extended to QRBMs. QRBMs enhance the generative capabilities of RBMs by leveraging quantum effects such as superposition and entanglement, enabling them to sample from high-dimensional and complex probability distributions efficiently. These properties make QRBMs particularly suitable for generating synthetic data in large-scale, complex applications such as anomaly detection and advanced generative tasks \cite{amin2018quantum}.

A key aspect of implementing QRBMs lies in mapping the RBM energy function to the Ising model \cite{bian2010ising}, a well-studied model in quantum mechanics and statistical physics. An energy function defines the Ising model:
\begin{equation}
H = -\sum_{i} h_i s_i - \sum_{i<j} J_{ij} s_i s_j,
\end{equation}
where $s_i \in \{-1, +1\}$ are spin variables, $h_i$ are local biases, and $J_{ij}$ are coupling coefficients between spins. The Ising model is the basis for quantum annealers, such as those developed by D-Wave Systems. The RBM energy function (Equation~\ref{eq:energy}) can be reformulated into an equivalent Ising Hamiltonian by encoding the visible and hidden units as spin variables, their biases as local fields, and their weights as couplings between spins. This mapping enables the direct implementation of RBM-like models on quantum hardware \cite{adachi2015application}.

Quantum annealing trains QRBM by finding low-energy configurations of the Ising Hamiltonian, which correspond to the most probable configurations in the RBM's probability distribution. This is particularly advantageous in high-dimensional and multimodal distributions, where classical methods often struggle due to local minima and high computational costs. By leveraging tunnelling, quantum annealers can escape local minima and explore the solution space more effectively. In this way, QRBMs are particularly suitable for generating synthetic data in applications like class imbalance correction, where they outperform classical oversampling techniques by better preserving the statistical properties of minority classes \cite{benedetti2016estimation}.
Advanced quantum architectures, such as D-Wave’s Pegasus topology, have enabled QRBMs to scale to larger problem sizes. The Pegasus topology provides improved qubit connectivity, allowing for more efficient embeddings of the Ising Hamiltonian corresponding to large RBMs. This advancement has opened new possibilities for tackling real-world problems, such as optimizing supply chains, enhancing cybersecurity systems, and advancing generative modelling for machine learning applications.

\section{Methodology}
\label{sec:method}
\begin{figure*}[h!]
    \centering
    \includegraphics[width=\textwidth]{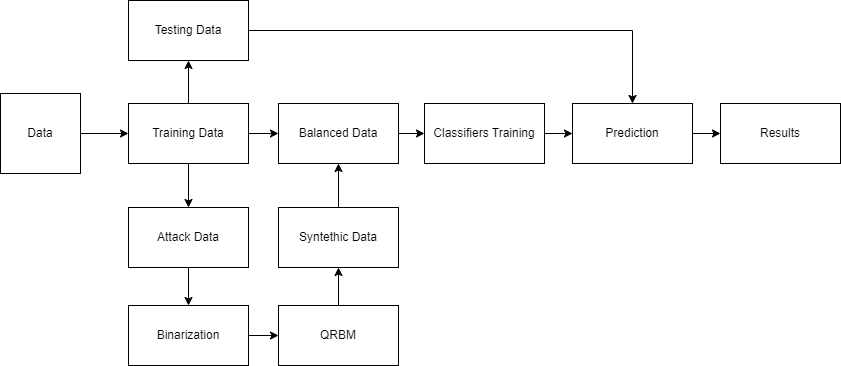} 
    \caption{Flow chart for balancing a dataset using QRBM.}
    \label{fig:Process-drawio}
\end{figure*}

Fig. \ref{fig:Process-drawio} illustrates the process for addressing the challenge of dataset imbalance using a QRBM. The workflow begins with the initial dataset, divided into two key components: training and testing data. 

The training data is split into attack datasets, and then the attack dataset undergoes a binarization process, converting the data into a binary format suitable for analysis by the QRBM. The QRBM is trained on the binarized attack data to generate synthetic data, which mirrors the underlying distribution and characteristics of the original attack data.

Combining synthetic data with the original training data produces a balanced dataset, ensuring that classes or data categories are equally represented. This newly balanced dataset is then used to train classifiers to identify patterns and make predictions based on the input data.

Once the classifiers are trained, their performance is evaluated using the testing data. The predictions generated by the classifiers are compared against the actual outcomes in the testing data to assess the model's accuracy, reliability, and overall effectiveness. The final step involves a comprehensive analysis of the prediction results, which provides valuable insights into the success of the balancing process and the performance of the classifiers.

\subsection{Dataset Preparation}
The study utilized the CICIDS2017 dataset, a comprehensive benchmark dataset developed by the Canadian Institute for Cybersecurity. CICIDS2017 is widely recognized for its relevance in evaluating IDS due to its diversity and realistic simulation of modern network traffic. The dataset captures a balanced representation of regular activity and various attack scenarios, providing a robust foundation for developing and testing advanced machine learning and quantum-based models \cite{sharafaldin2018toward}.

CICIDS2017 was created to address the shortcomings of earlier datasets, such as outdated attack types, lack of feature diversity, and unrealistic traffic generation. It includes over 3 million network flow records collected over 5 days, encompassing a mix of benign and malicious traffic. The malicious activities are categorized into six distinct attack types. The dataset is structured with over 83 features extracted using network flow analysis tools \cite{lashkari2017characterization}.  The feature richness makes CICIDS2017 an ideal dataset for testing anomaly detection systems and advanced generative modelling techniques.

Despite its comprehensiveness, the CICIDS2017 exhibits significant imbalance, with benign traffic far outnumbering malicious activity. Attack categories are severely underrepresented, necessitating synthetic data augmentation using QRBMs to balance the dataset.

\begin{table}[h]
\caption{Overall characteristics of CICIDS2017 dataset}
   \label{tab:ids_data_CICIDS2017}
   \begin{center}
   \def\arraystretch{1.2}
        \begin{tabular}{|c|c|}
        \hline
        Dataset Type                       & Multiclass \\ \hline
        Year of Release                    & 2017       \\ \hline
        Total Number of distinct instances & 2,830,540  \\ \hline
        Number of features                 & 83         \\ \hline
        Number of distinct classes         & 15         \\ \hline
        \end{tabular}
    \end{center}
\end{table}

\begin{table}[h]
\caption{Class instances CICIDS2017 dataset}
   \label{tab:ids_data_CICIDS2017_class_occurrence}
   \begin{center}
   \def\arraystretch{1.2}
    \begin{tabular}{|c|c|}
    \hline
    \textbf{Class Labels}    & \textbf{Number of instances} \\ \hline
    BENIGN                   & 2,359,087                    \\ \hline
    DoS Hulk                 & 231,072                      \\ \hline
    PortScan                 & 158,930                      \\ \hline
    DDoS                     & 41,835                       \\ \hline
    FTP-Patator              & 7,938                        \\ \hline
    SSH-Patator              & 5,897                        \\ \hline
    DoS slowloris            & 5,897                        \\ \hline
    DoS Slowhttptest         & 5,499                        \\ \hline
    Bot                      & 1,966                        \\ \hline
    Web Attack - Brute Force & 1,507                        \\ \hline
    Web Attack - XSS         &   652                        \\ \hline
    Infiltration             &    36                        \\ \hline
    Web Attack - Sql Injection &    21                        \\ \hline
    Heartbleed               &    11                        \\ \hline
    \end{tabular}
\end{center}
\end{table}

Table \ref{tab:ids_data_CICIDS2017} provides the overall characteristics of the CICIDS2017 dataset, while Table 
\ref{tab:ids_data_CICIDS2017_class_occurrence} provides an overview of the class distribution of the dataset. 

\subsection{QRBM Implementation on Pegasus Topology}
The Pegasus topology provides advanced qubit connectivity, with each qubit connecting to up to 15 others, making it ideal for embedding larger and more complex QRBMs \footnote{\url{https://docs.dwavesys.com/docs/latest/c_gs_4.html}}. This study developed an algorithm to implement large QRBMs (up to 172x120) on the Pegasus topology.

Our embedding algorithm (Algorithm \ref{alg:embedding_graph}) maps the visible and hidden units of the QRBM to physical qubits. It allows for shorter qubit chains, reducing the risk of chain breakage and improving the accuracy of the annealing process.  

Although the Pegasus architecture supports increased connectivity and enables larger embeddings, managing qubit allocation for increasingly complex QRBMs remains challenging. Preserving the bipartite structure of the QRBM and maintaining the integrity of interactions between visible and hidden units during physical embedding is critical. Our algorithm outperforms D-Wave's default embedding tool (minorminer.find\_embedding) as shown in Table \ref{tab:embedding_comparison_grouped}. In contrast, our algorithm efficiently identifies multiple QRBM configurations (ranging from 4x4 to 172x120) in milliseconds, eliminating the need for manual adjustments when the QRBM structure changes. Once the embedding is generated, the graph is passed to the annealer using the 
$J$ and $h$ matrices, ensuring seamless integration into the quantum computation process.

\begin{table}[h!]
\centering
\caption{Comparison of Embedding Algorithms for Different Configurations}
\label{tab:embedding_comparison_grouped}
\begin{tabular}{|c|c|c|c|}
\hline
\textbf{Algorithm}      & \textbf{Metric}         & \textbf{60x60} & \textbf{120x120} \\ \hline
\multirow{3}{*}{Our Algorithm} 
                        & Avg. Time (s)          & 0.012           & 0.014             \\ \cline{2-4} 
                        & Std. Deviation (s)     & 0.0012           & 0.0011             \\ \cline{2-4} 
                        & Avg. Chains with length $>$ 6                & 0              & 0               \\ \hline
\multirow{3}{*}{Minorminer} 
                        & Avg. Time (s)          & 214.94           & 960.02             \\ \cline{2-4} 
                        & Std. Deviation (s)     & 108.75           & 170.35             \\ \cline{2-4} 
                        & Avg.  Chains with length $>$ 6                 & 2              & 15               \\ \hline
\end{tabular}
\end{table}

Our algorithm begins by determining how the visible and hidden nodes are arranged on the processor. The number of visible and hidden nodes is divided into vertical and horizontal segments based on periodicity parameters, which dictate the spacing and structure of the node embeddings. Starting indices are defined for visible and hidden nodes based on the physical layout of the Quantum Processing Unit (QPU), guiding the placement of nodes within their respective layers. The mapping process involves iterating through these segments and assigning physical qubits to logical nodes. At the same time, couplings within each layer are identified and repeated across the QPU, ensuring that the intra-layer relationships are fully captured.

Once the nodes within each layer are mapped, the algorithm establishes inter-layer connections between visible and hidden nodes. This step pairs nodes from the two layers based on their logical indices and periodicity. These connections are stored as part of the coupling data and are later used to populate the coupling matrix. The algorithm's output includes the coupling matrix, visible and hidden node lists, and detailed intra-layer and inter-layer coupling records.

\subsection{Experimental Setup}

The experiments were conducted using a hybrid setup. A Dell R705X server with 128 logical cores and 503 GiB of RAM was utilized to perform classical machine learning tasks. The server ran Ubuntu 22.04 and Python 3.10.9 with Scikit-learn 1.5.2, and minorminer 0.2.16. SMOTE and RandomOverSampler were implemented to address the class imbalance. The QRBM balanced the datasets by generating 1,000 samples of attack class data in each anneal, with D-Wave's default annealing time of $20 \mu s$ and $124 \mu s$ for readout.

The evaluation metrics adopted were precision, recall, and F\textsubscript{1} scores, calculated across several classifiers: Support Vector Machines (SVM), Logistic Regression, Na\"{i}ve Bayes, Decision Trees, Gradient Boosting, K-Nearest Neighbors (KNN), and Random Forest.

\subsubsection{Data Preprocessing}
The preprocessing involved converting the dataset into a binary format compatible with the QRBM. Steps included normalization, encoding, and feature transformation to enhance the data's quality and compatibility.
Handling missing values was crucial in preprocessing the dataset, as unaddressed gaps could lead to biased or inaccurate models. Missing values in the dataset were addressed using mean and median imputation techniques, with invalid or incorrect entries removed to improve data integrity. In total, 1,305 NaN values and 1,310 infinite values were identified and eliminated from the dataset.
A correlation analysis was conducted to identify features with strong associations to address redundancy. Highly correlated features (absolute correlation $\geq 0.9$) and features with zero standard deviation were deemed redundant and removed. In total, 23 features with high correlations and eight with zero variance were removed, leaving 48 meaningful features. Duplicate records were identified and removed to prevent bias in the models. 307,078 duplicate entries were removed, leaving a final dataset of 2,524,847 records. This included 2,104,309 benign samples and 420,538 attack samples (Table \ref{tab:final_record}). 

Key features were extracted based on their significance in intrusion detection tasks and encoded into binary form using equation \ref{eq:binarization}:
\begin{equation}
\label{eq:binarization}
\text{Max(feature\_value)} - \text{Min(feature\_value)} \leq 2^N
\end{equation}
where $N$ is the number of bits required for encoding.

At the end of this process, we created a 120-bit vector, and we selected a QRBM with 120 visible and 120 hidden units.

\subsection{Balancing the Dataset using the QRBM}
The QRBM was trained using Algorithm \ref{alg:optimisation_quantum}, adapted from \cite{dixit2023quantum}, and the resulting parameters ($W$,$b$,$c$) were utilized in the embedding process via Algorithm \ref{alg:embedding_graph}. Each quantum annealing step produced 1,000 samples in milliseconds, with the visible unit states forming individual entries in the synthetic dataset. Repeating this process 1,700 times resulted in 1,683,771 synthetic attack samples, creating a perfectly balanced dataset of 4,208,618 samples. With each annealing process lasting only $20 \mu s$, we rapidly generated high-quality synthetic data.

\begin{table}[h!]
\centering
\caption{Train/Test Split}
\label{tab:final_record}
\begin{tabular}{|c|c|c|c|}
\hline
           & \textbf{Total} & \textbf{Benign} & \textbf{Attack} \\ \hline
\textbf{Training} & 1,767,393  & 1,473,016  & 294,377   \\ \hline
\textbf{Test}     & 757,454    & 631,293    & 126,161   \\ \hline
\textbf{Total}    & 2,524,847  & 2,104,309  & 420,538   \\ \hline
\end{tabular}
\end{table}

\section{Results}
\label{sec:results}

The results are tabulated and compared for four types of datasets: without balancing, balancing using SMOTE, RandomOversampler, and QRBM.

Table \ref{tab:time_to_balance} shows the time taken to balance the dataset using three different models: SMOTE, Random Over Sampler and QRBM. The SMOTE and Random Over Sampler methods took a relatively short time (0.75 and 0.23 seconds, respectively), and only 0.33 seconds for the QRBM. 

Table \ref{tab:Evaluation metrics without Balancing} shows the results of the evaluation metrics obtained when the dataset was trained without balancing the dataset. 

Table \ref{tab:Evaluation metrics after balancing using SMOTE} shows the results of the evaluation metrics obtained when the dataset was trained and modelled after balancing it using SMOTE.\\
Table \ref{tab:evaluation_oversampler} shows the results of the evaluation metrics obtained when the dataset was trained and modelled after balancing it using RandomOverSampler. 

Table \ref{tab:evaluation_RBM} shows the results of the evaluation metrics obtained when the dataset was trained and modelled after balancing it using QRBM.

\begin{table}[hbt!]
\centering
\centering
\caption{Time taken to balance the dataset}
\label{tab:time_to_balance}
\begin{tabular}{| c | c | c | c |c|}
\hline
  & SMOTE & RandomOverSampler & QRMB \\
\hline
Time(s) & 0.75 & 0.23 & 0.33 \\
\hline
\end{tabular}
\end{table}

\begin{table}[hbt!]
\centering
\caption{Evaluation metrics without Balancing}
\label{tab:Evaluation metrics without Balancing}
\begin{tabular}{| l | c | c | c |}
\hline
Model & Precision & Recall & F\textsubscript{1} Score\\
\hline
SVM & 24.67 & 25.81 & 24.84 \\
Na\"{i}ve Bayes & 51.81 & 47.27 & 27.04 \\
Logistic Regression & 24.74 & 24.95 & 24.84 \\
Gradient Boosting & 54.74 & 54.46 & 54.24 \\
KNN & 48.7 & 46.94 & 47.75 \\
Decision Tree & 50.35 & 51.17 & 50.41 \\
Random Forest & 77.3 & 50.97 & 53.29 \\
\hline

\end{tabular}

\end{table}

\begin{table}[hbt!]
\centering
\caption{Evaluation metrics after balancing using SMOTE}
\label{tab:Evaluation metrics after balancing using SMOTE}
\begin{tabular}{| l | c | c | c | }
\hline
Model & Precision & Recall & F\textsubscript{1} Score  \\
\hline
SVM & 74.8 & 63.63 & 57.69 \\
Na\"{i}ve Bayes & 70.18 & 70.32 & 64.14 \\
Logistic Regression & 77.93 & 70.88 & 64.84 \\
Gradient Boosting & 85.79 & 85.42 & 85.31 \\
KNN & 85.53 & 85.67 & 85.6 \\
Decision Tree & 86.73 & 86.7 & 86.5 \\
Random Forest & 89.49 & 89.51 & 89.5 \\
\hline

\end{tabular}

\end{table}

SVM shows a precision of 74.8\%, recall of 63.63\%, and an F\textsubscript{1} score of 57.69\%, indicating a balanced performance with substantial improvements compared to the unbalanced scenario. Na\"{i}ve Bayes also exhibits balanced metrics, with precision at 70.18\%, recall at 70.32\%, and an F\textsubscript{1} score of 64.14\%. Logistic Regression shows slightly higher metrics, with precision at 77.93\%, recall at 70.88\%, and an F\textsubscript{1} score of 64.84\%, demonstrating good model performance after balancing.

Gradient Boosting, KNN, and Random Forest stand out with significantly higher metrics. Gradient Boosting has precision, recall, and F\textsubscript{1} scores of approximately 85.4\%, while KNN shows similar consistency across the metrics with scores around 85.6\%. Random Forest leads with the highest precision at 89.49\%, recall at 89.51\%, and an F\textsubscript{1} score of 89.5\%, indicating superior performance in the balanced dataset.

Decision Tree also shows strong performance with precision at 86.73\%, recall at 86.7\%, and an F\textsubscript{1} score of 86.5\%, highlighting its effectiveness after SMOTE balancing.

\begin{table}[hbt!]
\centering
\caption{Evaluation metrics after balancing using RandomOverSampler}
\label{tab:evaluation_oversampler}
\begin{tabular}{| l | c | c | c |}
\hline
Model & Precision & Recall & F\textsubscript{1} Score \\
\hline
SVM & 74.86 & 63.55 & 57.57 \\
Na\"{i}ve Bayes & 67.24 & 69.69 & 63.05 \\
Logistic Regression & 78.2 & 70.74 & 64.63 \\
Gradient Boosting & 87.41 & 87.03 & 86.95 \\
KNN & 90.67 & 90.50 & 90.5 \\
Decision Tree & 96.08 & 95.98 & 95.99 \\
Random Forest & 96.05 & 95.95 & 95.96 \\
\hline

\end{tabular}

\end{table}


Table \ref{tab:evaluation_oversampler} shows that SVM reached a precision of 74.86\%, recall of 63.55\%, and an F\textsubscript{1} score of 57.57\%, indicating a balanced performance with substantial improvements compared to the unbalanced scenario. Na\"{i}ve Bayes also displays balanced metrics, with precision at 67.24\%, recall at 69.69\%, and an F\textsubscript{1} score of 63.05\%. Logistic Regression exhibits slightly higher metrics, with precision at 78.2\%, recall at 70.74\%, and an F\textsubscript{1} score of 64.63\%, demonstrating good model performance after balancing.

Gradient Boosting, KNN, and Random Forest stand out with significantly higher metrics. Gradient Boosting shows precision, recall, and F\textsubscript{1} scores around 87.03\% to 87.41\%, while KNN demonstrates high consistency across the metrics with scores around 90.5\%. Random Forest and Decision Tree lead with the highest metrics, where Decision Tree has a precision of 96.08\%, recall of 95.98\%, and an F\textsubscript{1} score of 95.99\%, while Random Forest follows closely with a precision of 96.05\%, recall of 95.95\%, and an F\textsubscript{1} score of 95.96\%, indicating superior performance in the balanced dataset.

\begin{table}[hbt!]
\centering
\caption{Evaluation metrics after balancing using QRBM}
\label{tab:evaluation_RBM}
\begin{tabular}{| l | c | c | c |}
\hline
Model & Precision & Recall & F\textsubscript{1} Score \\
\hline
SVM & 75.19 & 63.67 & 57.67 \\
Na\"{i}ve Bayes & 67.68 & 69.77 & 63.11 \\
Logistic Regression & 77.08 & 70.6 & 64.28 \\
Gradient Boosting & 87.94 & 87.54 & 87.46 \\
KNN & 91.15 & 90.93 & 90.91 \\
Decision Tree & 95.94 & 95.87 & 95.87 \\
Random Forest & 96.18 & 96.1 & 96.1 \\
\hline
\end{tabular}
\end{table}


Table \ref{tab:evaluation_RBM} illustrates various models' evaluation metrics (Precision, Recall, F\textsubscript{1} Score) after balancing using QRBM. The Random Forest achieves the highest precision at 96.18\%, and the Decision Tree follows closely with a precision of 95.94\%. KNN also performs well with a precision of 91.15\%. Other models like Gradient Boosting, Logistic Regression, Na\"{i}ve Bayes, and SVM have progressively lower precision values. Tree-based methods like Random Forests and Decision Trees demonstrate the best overall performance across all metrics. KNN also performs well but with slightly lower values. While effective, Gradient Boosting, Logistic Regression, Naive Bayes, and SVM do not achieve the same high precision, recall, and F\textsubscript{1} scores as the top-performing models.

\section{Discussion}
\label{sec:discussion}
QRBMs achieved superior performance compared to classical methods, such as SMOTE and RandomOverSampler, by generating high-quality synthetic samples that improved evaluation metrics like precision, recall, and F\textsubscript{1} scores across multiple ML models. Furthermore, QRBMs demonstrated remarkable time efficiency, completing the balancing process in just 0.33 seconds, as shown in Table \ref{tab:time_to_balance}. 

QRBMs exhibit clear advantages over classical methods such as SMOTE and RandomOverSampler by efficiently generating high-quality synthetic data using quantum annealing to model complex interactions. Unlike statistical approaches, QRBMs capture probabilistic dependencies across features, allowing for more nuanced sample generation. This capability is particularly critical in IDS applications, where the distribution of attack vectors is intricate and highly non-uniform, requiring sophisticated modelling techniques to ensure effective data balancing and robust detection.

The comparison of QRBMs with deep learning-based generative models, such as GANs, further highlights their unique advantages. While GANs can learn from data distributions and generate high-quality synthetic samples, they face challenges like mode collapse and demand significant computational resources, particularly for large-scale datasets. In contrast, QRBMs leverage quantum hardware to model energy landscapes directly, offering rapid sampling and scalability. Unlike GANs, QRBMs circumvent extensive training and hyperparameter tuning, making them a more efficient choice in specific scenarios. However, QRBMs have limitations, as quantum hardware restricts scalability due to limited qubit counts and noise. Additionally, hybrid quantum-classical workflows introduce complexities in preprocessing and postprocessing, requiring further advancements in hardware and algorithms to realize their full potential. Future research should include direct comparisons between QRBMs and GANs, focusing on computational efficiency, sample diversity, and downstream IDS performance.

QPU sampling time (134.92 $\mu s$) and readout time (94 $\mu s$) illustrate QRBM's capability to generate synthetic samples rapidly. Unlike classical methods, QRBMs maintain consistent runtimes across varying dataset sizes, making them well-suited for large-scale data preprocessing tasks.

QRBMs face challenges due to the limitations of current quantum hardware. Issues such as limited qubit connectivity, short coherence times, and error rates can impact their performance. Moreover, QRBM workflows' hybrid nature, combining quantum and classical computations, introduces additional data preprocessing and postprocessing complexity. The introduction of the Zephyr topology, with its enhanced qubit connectivity and density, has the potential to mitigate many of these hardware limitations, enabling the embedding of larger and more complex QRBMs while reducing chain breaks and improving sampling quality. Additionally, Zephyr's improved scalability and efficiency could streamline hybrid quantum-classical workflows by simplifying preprocessing and embedding steps, further unlocking the potential of QRBMs in real-world applications. 

Future research should focus on overcoming these challenges through advancements in quantum hardware and optimization techniques, in particular, developing more efficient embedding strategies to scale QRBM implementations for even larger and more complex datasets, exploring alternative quantum architectures and algorithms to optimize QRBM performance further and expanding the application of QRBMs in domains like healthcare, finance, and other real-time anomaly detection fields.

\section{Conclusion}
\label{sec:conclusion}

This study has demonstrated the potential of quantum-restricted Boltzmann Machines (QRBMs) as a powerful tool for addressing dataset imbalances e.g., in intrusion detection systems (IDS). By leveraging the advanced capabilities of the Pegasus architecture, we successfully implemented one of the largest QRBMs embedded on D-Wave’s quantum processor to date, with 120 visible and 120 hidden units. Notably, this scale of QRBM exceeds the capabilities of D-Wave’s default embedding tools, highlighting the significance of this achievement in generative modelling and quantum computing.

QRBMs generated high-quality synthetic data, significantly improving IDS performance across multiple metrics, including precision, recall, and F\textsubscript{1} scores. Compared to classical techniques such as SMOTE and RandomOverSampler, QRBMs provided superior results while maintaining remarkable computational efficiency. QRBMs' generative modelling capability, supported by the enhanced connectivity and flexibility of the Pegasus topology, enabled efficient sampling from complex energy landscapes, overcoming the limitations of classical approaches like Gibbs sampling. 

We therefore contribute by demonstrate the practical feasibility of large-scale QRBM implementations, pushing the boundaries of what can be achieved on existing quantum hardware.

\bibliographystyle{plain}
\bibliography{references}

\appendix
\section{Supplementary Information}

\subsection{Algorithms}
\begin{breakablealgorithm}
\caption{Embedding a QRBM on Pegasus}
\label{alg:embedding_graph}
\begin{algorithmic}[1]
\State \textbf{Input:} $n\_visible$, $n\_hidden$, $periodicity\_v$, $periodicity\_h$, $n\_periodicity$
\State \textbf{Output:} Logical-physical mappings for visible and hidden nodes, coupling matrix $J$

\State $H\_V \gets n\_visible / n\_periodicity$ 
\State $H\_H \gets n\_hidden / n\_periodicity$ 
\State Initialize $J\_coupling$, $visible\_nodes$, $hidden\_nodes$, and $J\_connections$ as empty lists

\State $startv \gets periodicity\_v$ 
\For{$z \gets 0$ to $H\_V - 1$}
    \For{$x \gets 0$ to $n\_periodicity - 1$}
        \State $n \gets startv + (n\_periodicity \cdot x) + (z \cdot periodicity\_v)$
        \State Append $n$ to $visible\_nodes$
        \For{$j \gets 0$ to $H\_H - 2$}
            \State Append $(n + j, n + j + 1)$ to $J\_coupling$
        \EndFor
    \EndFor
\EndFor

\State $starto \gets periodicity\_h$ 
\For{$z \gets 0$ to $H\_H - 1$}
    \For{$x \gets 0$ to $n\_periodicity - 1$}
        \State $p \gets starto + (n\_periodicity \cdot x) + (z \cdot periodicity\_h)$
        \State Append $p$ to $hidden\_nodes$
        \For{$j \gets 0$ to $H\_V - 2$}
            \State Append $(p + j, p + j + 1)$ to $J\_coupling$
        \EndFor
    \EndFor
\EndFor

\For{$x \gets 0$ to $H\_H - 1$}
    \For{$y \gets 0$ to $H\_V - 1$}
        \For{$t \gets 0$ to $n\_periodicity - 1$}
            \State $n \gets startv + (n\_periodicity \cdot t) + y \cdot periodicity\_v + x$
            \For{$k \gets 0$ to $n\_periodicity - 1$}
                \State $p \gets starto + (n\_periodicity \cdot k) + x \cdot periodicity\_h + y$
                \State Append $(n, p)$ to $J\_connections$
            \EndFor
        \EndFor
    \EndFor
\EndFor

\State $J \gets 0$
\For{Each $(a, b)$ in $J\_coupling$}
    \State $J[a, b] \gets -1$
\EndFor

\State \Return $J$, $visible\_nodes$, $hidden\_nodes$, $J\_coupling$, $J\_connections$
\end{algorithmic}
\end{breakablealgorithm}

\begin{breakablealgorithm}
\caption{Training of a QRBM using quantum annealing (based on \cite{dixit2023quantum})}\label{alg:optimisation_quantum}
\begin{algorithmic}[1]
\State $\epsilon \gets learning \, rate$ 
\State $b,c,W \gets random \, number$ 
\While{$not \,converged$}
    \State Sample a mini-batch of m examples $\{x^{(1)},...,x^{(m)}\}$ from the training set
    \State $V \gets \{x^{(1)},...,x^{(m)}\}$
    \State $H \gets \sigma(c+VW)$
    \State $\{h,J\} \gets \{b,c,W\}$
    \State $(V',H') \gets quantum\, annealing(h,J)$
    \State $W \gets W + \epsilon \left( \frac{V^TH}{m}-\frac{V'^TH'}{m_1} \right) $ 
    \State $b \gets b+ \epsilon \left( \frac{sum(V)}{m}-\frac{sum(V')}{m_1} \right) $ 
    \State $c \gets c+ \epsilon \left( \frac{sum(H)}{m}-\frac{sum(H')}{m_1} \right) $
    
\EndWhile
\end{algorithmic}
\end{breakablealgorithm}

\end{document}